\documentclass[twocolumn,tighten]{aastex61}
\hypersetup{linkcolor=cyan,citecolor=cyan,filecolor=cyan,urlcolor=cyan}
\submitjournal{ApJ Letters}

\usepackage{amsmath}
\usepackage{CJK}

\shorttitle{intrinsic shape of slow rotators}
\shortauthors{Hongyu~Li et al.}

\begin{document}
\begin{CJK*}{UTF8}{gbsn}

\title{SDSS-IV MaNGA: The intrinsic shape of slow rotator early-type galaxies}

\correspondingauthor{Hongyu Li}
\email{E-mail: hyli@nao.cas.cn}

\author[0000-0002-6518-9866]{Hongyu Li (李弘宇)}
\affiliation{National Astronomical Observatories, Chinese Academy of Sciences, 20A Datun Road, Chaoyang District, Beijing 100012, China}
\affiliation{University of Chinese Academy of Sciences, Beijing 100049, China}

\author[0000-0001-8317-2788]{Shude Mao}
\affiliation{Physics Department and Tsinghua Centre for Astrophysics, Tsinghua University, Beijing 100084, China}
\affiliation{National Astronomical Observatories, Chinese Academy of Sciences, 20A Datun Road, Chaoyang District, Beijing 100012, China}
\affiliation{Jodrell Bank Centre for Astrophysics, School of Physics and Astronomy, The University of Manchester, Oxford Road, Manchester M13 9PL, UK}

\author[0000-0002-1283-8420]{Michele Cappellari}
\affiliation{Sub-Department of Astrophysics, Department of Physics, University of Oxford, Denys Wilkinson Building, Keble Road, Oxford, OX1 3RH, UK}

\author[0000-0002-4901-0571]{Mark T. Graham}
\affiliation{Sub-Department of Astrophysics, Department of Physics, University of Oxford, Denys Wilkinson Building, Keble Road, Oxford, OX1 3RH, UK}

\author[0000-0002-6155-7166]{Eric Emsellem}
\affiliation{Universit\'e Lyon 1, Observatoire de Lyon, Centre de Recherche Astrophysique de Lyon and Ecole Normale Sup\'erieure de Lyon, 9 avenue 
Charles Andr\'e, F-69230 Saint-Genis Laval, France}
\affiliation{European Southern Observatory, Karl-Schwarzschild-Str. 2, 85748 Garching, Germany}

\author{R.~J.~Long}
\affiliation{National Astronomical Observatories, Chinese Academy of Sciences, 20A Datun Road, Chaoyang District, Beijing 100012, China}
\affiliation{Jodrell Bank Centre for Astrophysics, School of Physics and Astronomy, The University of Manchester, Oxford Road, Manchester M13 9PL, UK}



\begin{abstract}
By inverting the distributions of galaxies' apparent ellipticities and misalignment angles 
(measured around the projected half-light radius $R_{\rm e}$) between their photometric and kinematic axes,
we study the intrinsic shape distribution of 189 slow rotator early-type galaxies with stellar masses 
$2\times 10^{11} M_{\odot}<M_\ast<2\times 10^{12} M_{\odot}$, extracted from a sample of about 2200 galaxies
with integral-field stellar kinematics from the DR14 of the SDSS-IV MaNGA IFU survey. Thanks to the large 
sample of slow rotators, Graham+18 showed that there is clear structure in the misalignment angle distribution,
with two peaks at both $0^{\circ}$ and $90^{\circ}$ misalignment (characteristic of oblate and prolate
rotation respectively).
Here we invert the observed distribution from Graham+18. The large sample allows us to go beyond the known
 fact that slow rotators are
weakly triaxial and to place useful constraints on their intrinsic triaxiality distribution (around $1R_{\rm e}$)
for the first time. The shape inversion is generally non-unique. However, we find that, for a wide set of model 
assumptions, the observed distribution clearly requires a dominant triaxial-oblate population. For some of our models,
the data suggest a hint for a minor triaxial-prolate population, but a dominant prolate population is ruled out.
\end{abstract}




\section{Introduction} \label{sec:intro}
The intrinsic shape distribution of early-type galaxies has been studied for decades by statistically
inverting the distribution of their ellipticities
(e.g. \citealt{Sandage1970,Binney1981,Fasano1991,Lambas1992,Ryden1992,Kimm2007}) or the distribution
of their ellipticities and misalignment angles between photometric and kinematic axes 
(e.g. \citealt{Weijmans2014,Foster2017}). Although the recovered intrinsic shape distributions from such
methods are in general non-unique \citep{Franx1991}, the fact that all the fast rotators have kinematic
axes aligned with their photometric axes \citep{Cappellari2007,Emsellem2007,Krajnovic2011,Fogarty2015}
can only be explained if, as a class, they are nearly axisymmetric (with or without bars).
For slow rotators, however, the two axes can be misaligned. Previous studies show that
slow rotators are weakly triaxial \citep{Weijmans2014,Foster2017}, but it is difficult to see any significant
features of the misalignment angle distribution due to their limited sample sizes. 
A more detailed review on this topic can be found in section 3.3 of \citet{Cappellari2016}.

In recent cosmological hydrodynamic simulations (e.g. fig.~7 of \citealt{Schaller2015}, fig.10 of
\citealt{Li2016},\citealt{Velliscig2015} and \citealt{Li2018}), many massive galaxies are found to
be prolate-like, manifesting as mostly slow rotators with different kinematic
misalignment angles and formed by major mergers with radial orbits \citep{Li2018}. This theory is broadly consistent
with the observation that genuine kinematic misalignment between photometry and kinematics, including some
nearly 90-degree misalignments \citep{Krajnovic2011}, only happens in slow rotators, which are found
above a characteristic stellar mass ($M_{\rm crit}\approx2\times10^{11}M_{\odot}$;
e.g. fig. 11 of \citealt{Emsellem2011}; \citealt{Cappellari2013a}; \citealt{Cappellari2016}).
It would be useful to study the intrinsic shape distribution of slow rotators observationally and put
constraints on galaxy formation models.

Some recent works have interpreted the existence of a handful of massive
galaxies with 90-degrees misalignment as circumstantial evidence for the existence of prolate
galaxies, e.g. in the CALIFA survey \citet{Tsatsi2017} and the MUSE Most Massive Galaxies
(M3G) survey \citep{Krajnovic2018}. Especially in the M3G survey, half of 
the massive ($M>10^{12} M_{\odot}$) galaxies have 90-degrees misalignments. 
However a 90-degree misalignment is also
naturally expected as a significant fraction for triaxial galaxies (see e.g. this paper).
This implies that, unless all slow rotators were 90-degree misaligned, which is not the case
(see fig.6 of \citealt{Cappellari2016}), one cannot interpret 90-degree misalignments as
evidence for prolate galaxies. Conclusions on galaxy shapes cannot be reached without a
statistical study but the number statistics are so far too limited for this kind of analysis.

MaNGA \citep{Bundy2015} is currently the largest IFU survey observed by the $\rm 2.5\, m$ Sloan Telescope 
\citep{Gunn2006}, and as reported by \citet{Graham2018}, provides a sufficiently large number of slow rotators
to detect actual structure in the distribution of kinematic misalignments.
In this Letter we analyze the distributions of ellipticity and misalignment angle measured around the effective
radius to provide the first statistical study of the shape distribution of slow rotators, and goes 
beyond the well-established fact that they are ``weakly triaxial" (e.g. review by \citealt{Cappellari2016}).

The structure of this Letter is as follows. In Section~\ref{sec:method}, we introduce
the galaxy sample and the methods we use. In Section \ref{sec:rst}, we show the
results. In Section \ref{sec:conclusion}, we summarize our results and draw conclusions.

\section{Sample and Methods}
\label{sec:method}
\subsection{Sample selection and data}
\label{sec:data}
We use the galaxy sample from \citet{Graham2018}, based on SDSS-IV DR14 MaNGA
\citep{DR14}, which has $\sim 2700$ galaxies. The measured ellipticity $\varepsilon$,
misalignment angle $\Psi_{\rm mis}$
(around the effective radius),  beam-corrected angular momentum parameter $\lambda_{\rm R_e}$ and 
kinematic classification are from table~2 of \citet{Graham2018}. We select massive slow
rotators using two alternative techniques. The first is quantitative: we select galaxies with 
$\lambda_{\rm R_e}<0.08+\varepsilon/4$ and $\varepsilon<0.4$ \citep{Cappellari2016}, 
as indicated by the black solid lines in fig.~11 of \citet{Graham2018}. The second one is 
qualitative: we select galaxies with visually-classified ETGs as non-regular rotators.
These galaxies do not exhibit a spider diagram velocity map and this is thought to be an
indicator of a slow rotator (see \citealt{Graham2018} and fig. 4 of \citealt{Cappellari2016} for more details). 
In both samples, we remove galaxies with $M\equiv 2\times M_{1/2}<2\times 10^{11} M_{\odot}$, where
$M_{1/2}$ is the dynamical mass within the 3-dimensional half-light radius ($r_{1/2}$) from
table~1 of \citet{Li2018b}. When the dark matter is small within the effective radius, 
$M$ defined above is a good proxy of the stellar mass \citep{Cappellari2013b}.
This cut on mass is needed because most slow or non-regular rotators with 
$M<2\times 10^{11} M_{\odot}$ are not genuine slow rotators \citep{Graham2018}.
Finally, we have 189 galaxies in the slow rotator sample and 74 galaxies in the 
non-regular sample. We note here that 72 of the 74 non-regular rotators are also in
the 189 slow rotator sample.

\begin{figure*}
\includegraphics[width=0.95\textwidth]{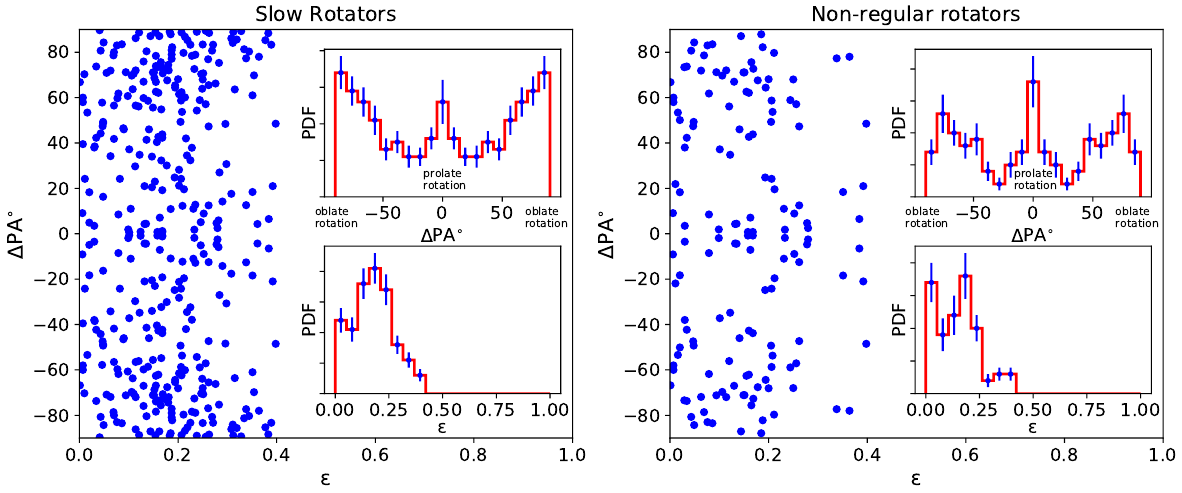}
\caption{The observed distribution of ellipticity $\varepsilon$ and kinematic misalignment $\Delta PA$
         for slow rotators (left) and non-regular rotators (right). For clearer visualisation, we shift
         the kinematic misalignment angle $\psi_{\rm mis}$ by $90^{\circ}$ and symmetrize their values
         about $0^{\circ}$, i.e. $|\Delta PA|=90^{\circ}-|\psi_{\rm mis}|$. By doing this, the peak around
         the prolate rotation can be clearly seen at the centre of the distribution, 
         i.e. $\Delta PA=0^{\circ}$. 
         In each panel, every blue dot represents an observed galaxy, the red histograms show their 1D 
         distribution, and the blue error bars show the error estimated using the bootstrapping method.
	     } 
\label{fig:obs}
\end{figure*}
In Fig.~\ref{fig:obs}, we show the 1D histograms and the 2D distribution of the observed ellipticity 
$\varepsilon$ and kinematic misalignment $\psi_{\rm mis}$ for slow rotators and non-regular rotators
from \citet{Graham2018}. We shift and symmetrize the misalignment
angle distribution. We note that we did not modify the data values in the symmetrization but just plot the data twice between $[-90^{\circ},0^{\circ}]$ and $[0^{\circ},90^{\circ}]$ for clearer visualisation.
In the model fitting, we still use the original distribution as input. 
For normal axisymmetric rotation, $\Delta PA=\pm 90^{\circ}$. The uncertainties estimated by the
bootstrapping method are shown by the blue error bars. As one can see, the peak around $\Delta PA=0^{\circ}$
has a confidence level more than two-sigma.

\subsection{Projection of luminosity density and velocity field}
\label{sec:theory}
The projection of the luminosity density of a traxial galaxy has been discussed in 
\citet{Stark1977}, \citet{Binney1985} and \citet{Franx1988}. Here we just make a
brief summary. Following \citet{Binney1985}, we assume the luminosity density of
a triaxial galaxy could be described by similar coaxial ellipsoids, i.e. 
$\rho=\rho(m^2)$, where $m^2=x^2/a^2+y^2/b^2+z^2/c^2$, and $x-$, $y-$ and $z-$axes are
aligned with the principal axes with $a>b>c$. The axis ratios are defined as
\begin{equation}
	\zeta=b/a, \qquad
    \xi=c/a, \qquad
    \eta=c/b.     
\label{eq:axis-ratio}
\end{equation}
When a galaxy is viewed along the direction $(\theta, \phi)$ in polar coordinates,
the projected ellipticity $\varepsilon$ and the minor axes position angle $\Gamma_{\rm minor}$
with respect to the projected $z-$axis are completely determined by the intrinsic axes ratios
(equations 12 and 13 in \citealt{Binney1985})
\begin{equation}
\varepsilon(\theta, \phi; \xi, \zeta) = 1-\sqrt{\frac{A+C-\sqrt{(A-C)^2+B^2}}{A+C+\sqrt{(A-C)^2+B^2}}},
\label{eq:eps}
\end{equation}
\begin{equation}
\Gamma_{\rm minor}(\theta, \phi; \xi, \zeta) = \frac{1}{2} \arctan \left(\frac{B}{A-C}\right),
\label{eq:Gamma_minor}
\end{equation}
with 
\begin{equation}
A\equiv \frac{\cos^2\theta}{\xi^2}\left(\sin^2\phi + \frac{\cos^2\phi}{\zeta^2}\right)+\frac{\sin^2\theta}{\zeta^2},
\label{eq:A}
\end{equation}
\begin{equation}
B\equiv \cos\theta\sin 2\phi\left(1-\frac{1}{\zeta^2}\right) \frac{1}{\xi^2},
\label{eq:B}
\end{equation}
\begin{equation}
C\equiv \left(\frac{\sin^2\phi}{\zeta^2}+\cos^2\phi\right)\frac{1}{\xi^2}.
\label{eq:C}
\end{equation}

In a triaxial galaxy, the angular momentum vector is not required to be aligned with the
principal axes, but can lie anywhere in the $x-z$ plane \citep{Statler1987}. Following 
\citet{Franx1991}, we define $0^{\circ}\le \psi_{\rm int}\le 90^{\circ}$ as the angle between
the $z$-axis and the angular momentum vector. The position angle of the apparent angular 
momentum $\Gamma_{\rm kin}$ with respect to the projected $z-$axis only depends on the viewing
angle and the intrinsic misalignment (equation 6 in \citealt{Franx1991})
\begin{equation}
\tan\Gamma_{\rm kin}=\frac{\sin\phi\sin\psi_{\rm int}}{-\cos\phi\cos\theta\sin\psi_{\rm int}+\sin\theta\cos\psi_{\rm int}}.
\label{eq:Gamma_kin}
\end{equation}
The kinematic misalignment angle $\Psi$ between the projected minor axis and the projected
angular momentum is given by
\begin{equation}
\sin\Psi=|\sin(\Gamma_{\rm kin}-\Gamma_{\rm minor})|, \quad 0^{\circ}\le \Psi \le 90^{\circ},
\label{eq:Psai}
\end{equation}
where $\Psi$ corresponds to our observed quantity $|\psi_{\rm mis}|$ from \citet{Graham2018}.

\subsection{Monte Carlo simulations}
\label{sec:MC}
From Section~\ref{sec:theory}, we know the apparent kinematic misalignment angle $\Psi$ and
apparent ellipticity $\varepsilon$ are determined by the intrinsic axis ratios ($\zeta$, $\xi$),
intrinsic misalignment angle of the angular momentum $\psi_{\rm int}$ and  viewing angle
($\theta$, $\phi$), i.e. $\Psi=\Psi(\zeta,\xi,\psi_{\rm int},\theta,\phi)$,
$\varepsilon=\varepsilon(\zeta,\xi,\theta,\phi)$.
If we assume the viewing angle is random (i.e. flat distributions in $[-1, 1]$ for
$\cos\theta$ and $[0, 2\pi]$ for $\phi$),
and the distribution of the axis ratios and intrinsic misalignment of the galaxies in our sample
can be described by some distribution functions, e.g. $p(\zeta)$, 
$p(\xi)=p({\xi}/{\zeta}|\zeta)p(\zeta)=p(\eta|\zeta)p(\zeta)$ and
$p(\psi_{\rm int})$, we can calculate the model probability distribution of $\Psi$ and $\varepsilon$
\begin{equation}
p(\Psi,\varepsilon)=p[\Psi,\varepsilon|p(\zeta),p(\eta),p(\psi_{\rm int})]
\label{eq:prob}
\end{equation}
The reason for using $p(\eta)$ instead of $p(\xi)$ is because $\eta$ is between 0 and 1
and is independent of $\zeta$, which makes it easier and faster to implement numerically,
while $\xi$ has to be smaller than $\zeta$. We also confirmed by visual examination that the distribution
$p(\xi)$ derived from $p(\zeta)$ and $p(\eta)$ in our model is physically reasonable and 
can cover various shapes of distributions.

We use two different models to parameterize $p(\zeta)$, $p(\eta)$ and $p(\psi_{\rm int})$. The first
one is a single-population model, in which we assume $p(\zeta)$, $p(\eta)$ and $p(\psi_{\rm int})$
are independent and can be described by single truncated Gaussian functions
\begin{equation}
p(\zeta|\mu_{\zeta}, \sigma_{\zeta})\propto \frac{1}{\sqrt{2\pi\sigma_{\zeta}^2}}\exp\left(\frac{(\zeta-\mu_{\zeta})^2}{\sigma_{\zeta}^2}\right), \quad 0\le\zeta\le 1
\label{eq:P_zeta}
\end{equation}
\begin{equation}
p(\eta|\mu_{\eta}, \sigma_{\eta})\propto \frac{1}{\sqrt{2\pi\sigma_{\eta}^2}}\exp\left(\frac{(\eta-\mu_{\eta})^2}{\sigma_{\eta}^2}\right), \quad 0\le\eta\le 1
\label{eq:P_eta}
\end{equation}
\begin{equation}
\begin{split}
& p(\psi_{\rm int}|\mu_{\psi_{\rm int}}, \sigma_{\psi_{\rm int}})\propto \\
& \qquad \frac{1}{\sqrt{2\pi\sigma_{\psi_{\rm int}}^2}}\exp\left(\frac{(\psi_{\rm int}-\mu_{\psi_{\rm int}})^2}{\sigma_{\psi_{\rm int}}^2}\right), \quad 0\le\psi_{\rm int}\le \frac{\pi}{2}
\label{eq:P_psi_int}
\end{split}
\end{equation}
The probability outside the boundary is truncated and set to 0. 
There are 6 free parameters for a single-population model, they are $\mu_{\zeta}$, $\sigma_{\zeta}$,
$\mu_{\eta}$, $\sigma_{\eta}$, $\mu_{\psi_{\rm int}}$ and $\sigma_{\psi_{\rm int}}$.

The second model is a two-population model, in which we assume $p(\zeta)$, $p(\eta)$ and 
$p(\psi_{\rm int})$ are independent and can be described by two truncated Gaussian functions,
with each Gaussian representing a galaxy population (e.g. oblate, triaxial or prolate)
\begin{equation}
\begin{split}
& p(\zeta|\mu_{\zeta_1}, \sigma_{\zeta_1}, \mu_{\zeta_2}, \sigma_{\zeta_2})= \\
& \qquad f_1p(\zeta|\mu_{\zeta_1}, \sigma_{\zeta_1}) + (1-f_1) p(\zeta|\mu_{\zeta_2}, \sigma_{\zeta_2}),
\label{eq:P_zeta2}
\end{split}
\end{equation}
\begin{equation}
\begin{split}
& p(\eta|\mu_{\eta_1}, \sigma_{\eta_1}, \mu_{\eta_2}, \sigma_{\eta_2})= \\
& \qquad f_1p(\eta|\mu_{\eta_1}, \sigma_{\eta_1}) + (1-f_1) p(\eta|\mu_{\eta_2}, \sigma_{\eta_2}),
\label{eq:P_eta2}
\end{split}
\end{equation}
\begin{equation}
\begin{split}
& p(\psi_{\rm int}|\mu_{\psi_{\rm int_1}}, \sigma_{\psi_{\rm int_1}}, \mu_{\psi_{\rm int_2}}, \sigma_{\psi_{\rm int_2}}) ={} \\
& \qquad f_1p(\psi_{\rm int}|\mu_{\psi_{\rm int_1}}, \sigma_{\psi_{\rm int_1}})+(1-f_1)p(\psi_{\rm int}|\mu_{\psi_{\rm int_2}}, \sigma_{\psi_{\rm int_2}}).
\label{eq:P_psi_int2}
\end{split}
\end{equation}
where $f_1$ represents the fraction of the galaxies in the first galaxy population. The functional form and
truncation boundary for each population are exactly the same as in equations~\ref{eq:P_zeta},
\ref{eq:P_eta} and \ref{eq:P_psi_int}. There are 13 free parameters in this model, they are $\mu_{\zeta_1}$,
$\sigma_{\zeta_1}$, $\mu_{\eta_1}$, $\sigma_{\eta_1}$, $\mu_{\psi_{\rm int_1}}$, $\sigma_{\psi_{\rm int_1}}$,
$\mu_{\zeta_2}$, $\sigma_{\zeta_2}$, $\mu_{\eta_2}$, $\sigma_{\eta_2}$, $\mu_{\psi_{\rm int_2}}$,
$\sigma_{\psi_{\rm int_2}}$ and $f_1$. 

In the models above, we assume the intrinsic misalignment is independent of the intrinsic shape. However,
a one-to-one relation is assumed in previous studies \citep{Weijmans2014,Foster2017}
\begin{equation}
\label{eq:shape-misalignment}
\tan \psi_{\rm int}=\sqrt{\frac{T}{1-T}},
\end{equation}
where $T=\frac{a^2-b^2}{a^2-c^2}$ is the triaxial parameter (i.e. the intrinsic shape). 
We also adapt this one-to-one relation in our single- and two-population models in our Monte 
Carlo simulations. In summary, the models we use are
\begin{enumerate}
	\item a single-population model with independent intrinsic misalignment, 6 free parameters;
    \item a two-population model with independent intrinsic misalignment, 13 free parameters;
    \item a single-population model with intrinsic misalignment following the model of 
    	  \citet{Weijmans2014}, 4 free parameters;
    \item a two-population model with intrinsic misalignment following the model of 
    	  \citet{Weijmans2014}, 9 free parameters.
\end{enumerate}


For all the models, we estimate $p(\Psi, \varepsilon)$ in 
equation~\ref{eq:prob} numerically. With a given set of model parameters $\vec{p}$, we first sample 
$n$ points of $\cos \theta$, $\phi$ from a flat distribution, and $\zeta$, $\eta$ and $\psi_{\rm int}$
from $p(\zeta|\vec{p})$, $p(\eta|\vec{p})$ and $p(\psi_{\rm int}|\vec{p})$, respectively (for the
Weijman+14 models, $\psi_{\rm int}$ is calculated directly using equation~\ref{eq:shape-misalignment}).
We then calculate the model predicted $\varepsilon$ and $\Psi$ for every point using 
equation~\ref{eq:eps} and \ref{eq:Psai} respectively, and bin all the points on a 2-dimensional
histogram (25 by 25). Finally, we normalise the histogram, and use bilinear interpolation to 
calculate the probability for a given $(\Psi, \varepsilon)$. We set $n=9,000,000$ in our calculation,
which is enough to produce a smooth probability distribution.

We perform the inversion, i.e. infer $p(\zeta)$, $p(\xi)$ and $p(\psi_{\rm int})$ from the observed
distribution, by maximizing the likelihood of the observed distribution, defined as 
\begin{equation}
\ln L = \sum_{i}\ln p(|\psi_{\rm mis}^i|, \varepsilon_i) 
\label{eq:maximum_likelihood}
\end{equation}
where $\psi_{\rm mis}^i$ and $\varepsilon_i$ are the apparent kinematic misalignment angle and ellipticity
for the $i$-th observed galaxy. $p(\Psi, \varepsilon)$ is defined in equation~\ref{eq:prob} and estimated 
numerically using the method described above. The sum is over all the galaxies in the sample (i.e. slow
rotators or non-regular rotators described in Section~\ref{sec:data}). We use a python implementation
of the MCMC algorithm ({\bf emcee}, \citealt{Foreman2013}) to obtain the parameters which maximize the
likelihood as well as the probability distribution of the model parameters. We use 200 walkers and run
for 3000 and 12000 steps for the single-population model and two-population model, respectively. 
In addition to {\bf emcee}, we also tried another MCMC package {\bf AdaMet} \citep{Cappellari2013b}. Those two
packages give similar results. The parameters are sampled within their mathematical limits. For the 
single-population model, they are:
$[0, 1]$ for $\mu_{\zeta}$ and $\mu_{\eta}$;
$[0, 1]$ for $\sigma_{\zeta}$ and $\sigma_{\eta}$;
$[0, \pi/2]$ for $\mu_{\psi_{\rm int}}$;
$[0, \frac{3}{2}\pi]$ for $\sigma_{\psi_{\rm int}}$.
For the two-population model, the boundaries are
$[0, 1]$ for $\mu_{\zeta_1}$, $\mu_{\eta_1}$, $\mu_{\zeta_2}$ and $\mu_{\eta_2}$;
$[0, 1]$ for $\sigma_{\zeta_1}$, $\sigma_{\eta_1}$, $\sigma_{\zeta_2}$ and $\sigma_{\eta_2}$;
$[0, \pi/2]$ for $\mu_{\psi_{\rm int_1}}$ and $\mu_{\psi_{\rm int_2}}$;
$[0, \pi/2]$ for $\sigma_{\psi_{\rm int_1}}$ and $\sigma_{\psi_{\rm int_2}}$;
$[0, 1]$ for $f_1$.  

\section{Results}
\label{sec:rst} 
In this section, we show the fitting results of the observed distribution for all the models.
To test the robustness of our results, we fit both the slow rotator
samples selected with different techniques in Section~\ref{sec:data}.
The fitting is performed using the methods described in Section~\ref{sec:MC}.

\begin{figure*}
\includegraphics[width=0.97\textwidth]{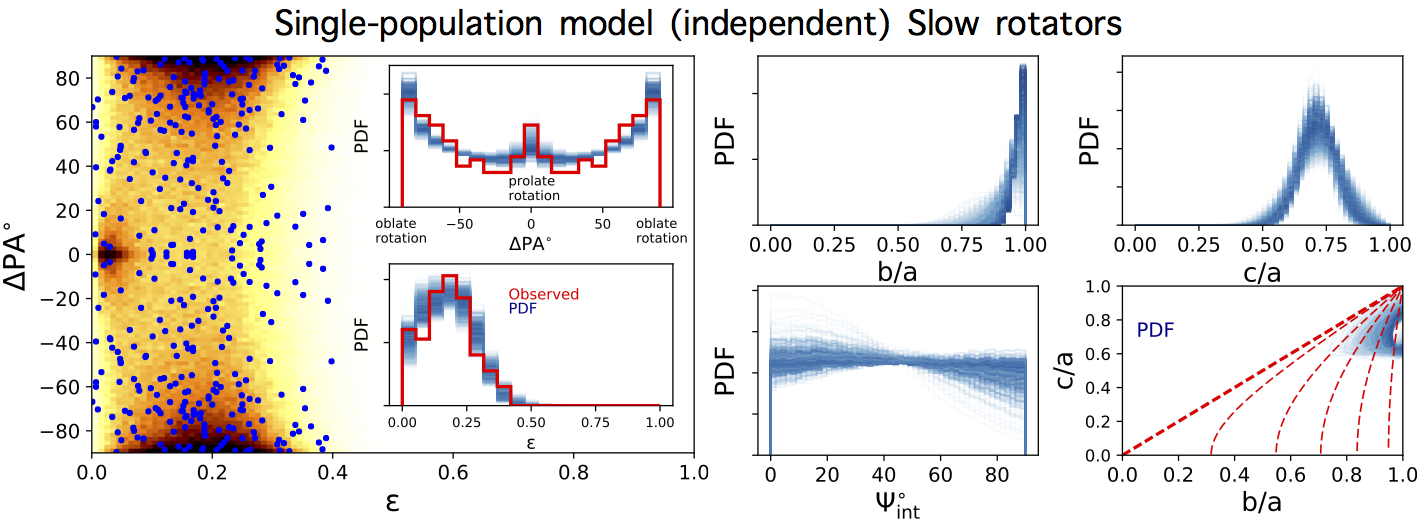}
\includegraphics[width=0.97\textwidth]{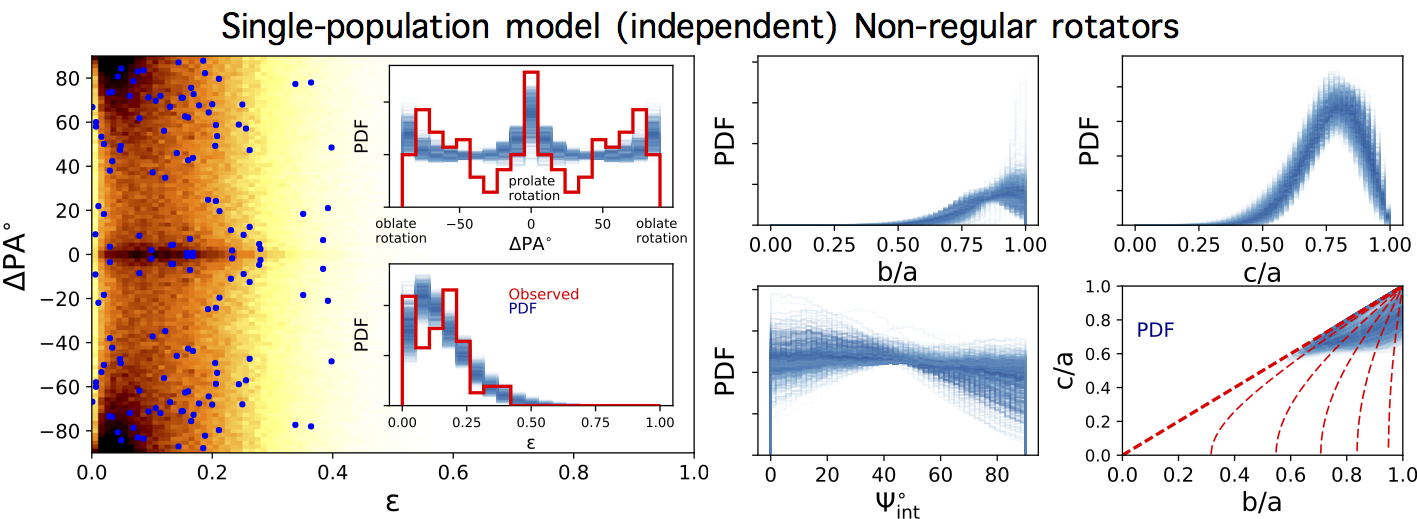}
\caption{Single-population models with independent intrinsic misalignment for massive slow rotators
         classified using two alternative 
         techniques, namely either (i) from the $(\lambda_{\rm R_e}, \varepsilon)$ diagram (top)
         or (ii) visually classified from their kinematic maps (bottom). In each
		 left panel, the blue dots show the observed distribution, and the colour map shows the
         model distribution which maximizes the likelihood. The red histograms show the
         observed distribution. The blue lines show the probability
         distribution, which are randomly selected from the MCMC chain and colour coded by their 
         likelihood. The bluer the colour, the larger the likelihood. We use the same method described
         in Fig.~\ref{fig:obs} to shift and symmetrize the distribution for visualisation purposes only.
         We note that we still use the original distribution in the model fitting process. 
         The predicted axis ratio and intrinsic misalignment are shown by the blue lines in the
         right panels.
         In the b/a vs. c/a panel (lower right), every blue
         line shows the 1$\sigma$ ($68\%$) contour of the 2D axis ratio distribution of that
         model. The red thin-dashed lines, from left to right, show the line of constant triaxial
         parameter $T=\frac{a^2-b^2}{a^2-c^2}$: 0.1, 0.3, 0.5, 0.7 and 0.9. The red thick-dashed line
         shows the position where the axis ratio $b/a=c/a$.
         } 
\label{fig:single}
\end{figure*}

\begin{figure*}
\includegraphics[width=0.97\textwidth]{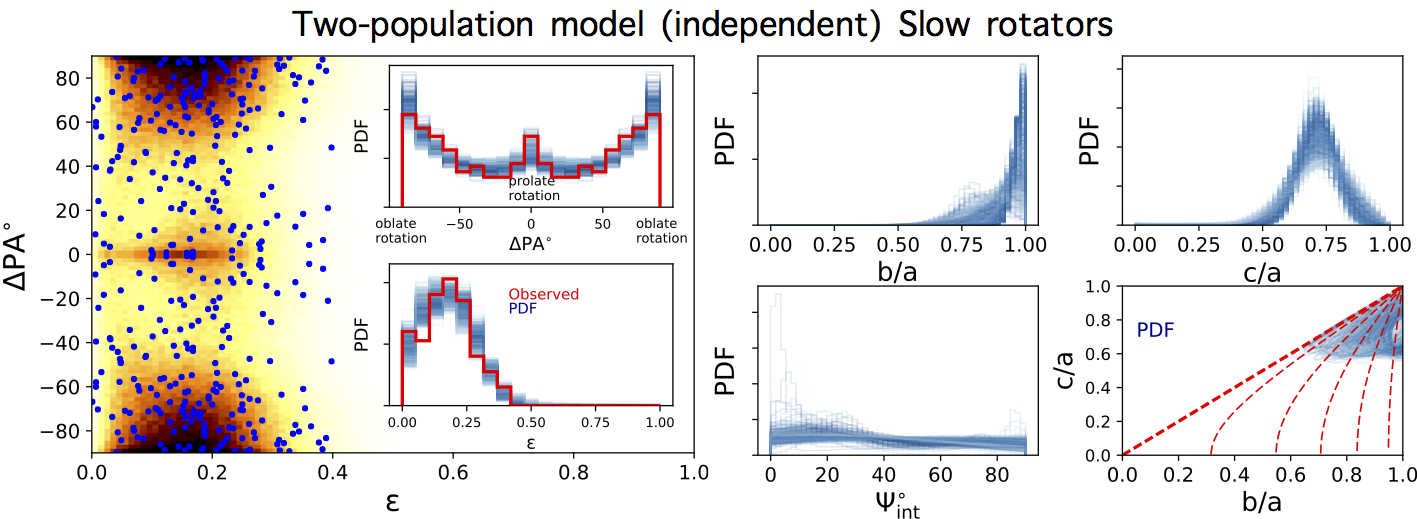}
\includegraphics[width=0.97\textwidth]{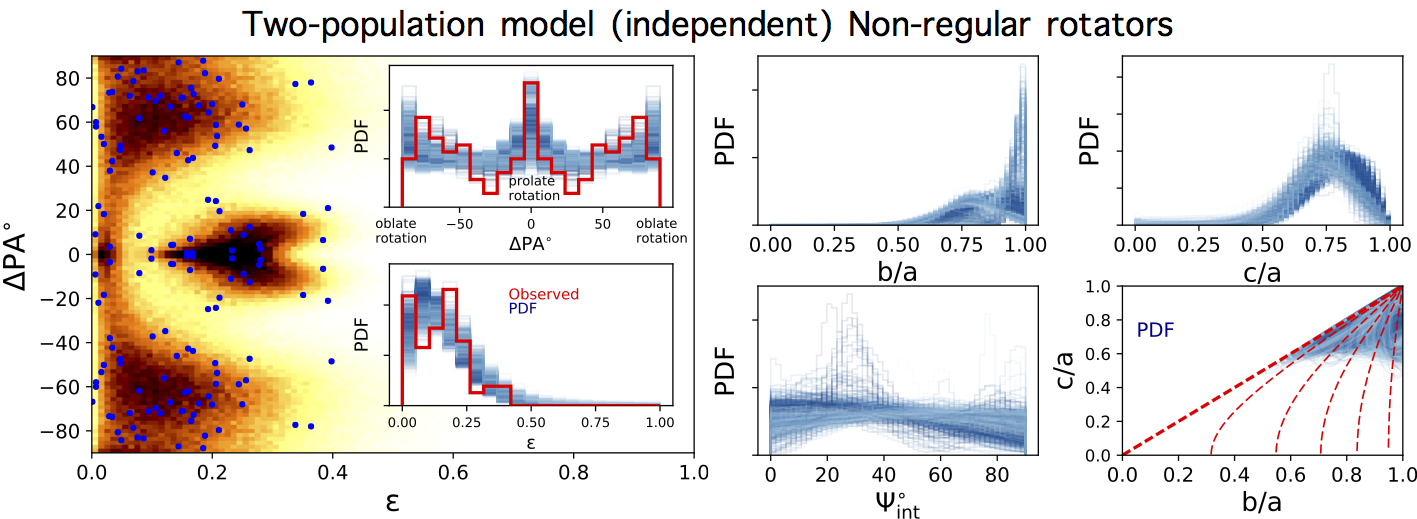}
\caption{Two-population model with independent intrinsic misalignment for slow rotators
         (top) and non-regular rotators (bottom). The other labels and
         legends are the same as Fig.~\ref{fig:single}.}
\label{fig:double}
\end{figure*}

\begin{figure*}
\includegraphics[width=0.97\textwidth]{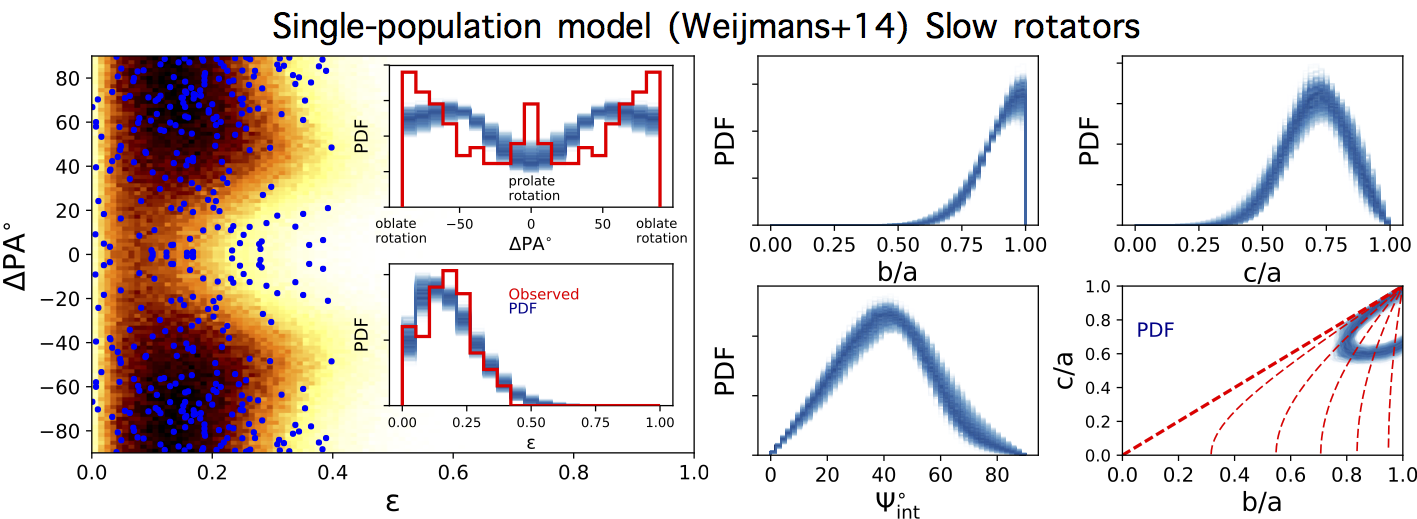}
\includegraphics[width=0.97\textwidth]{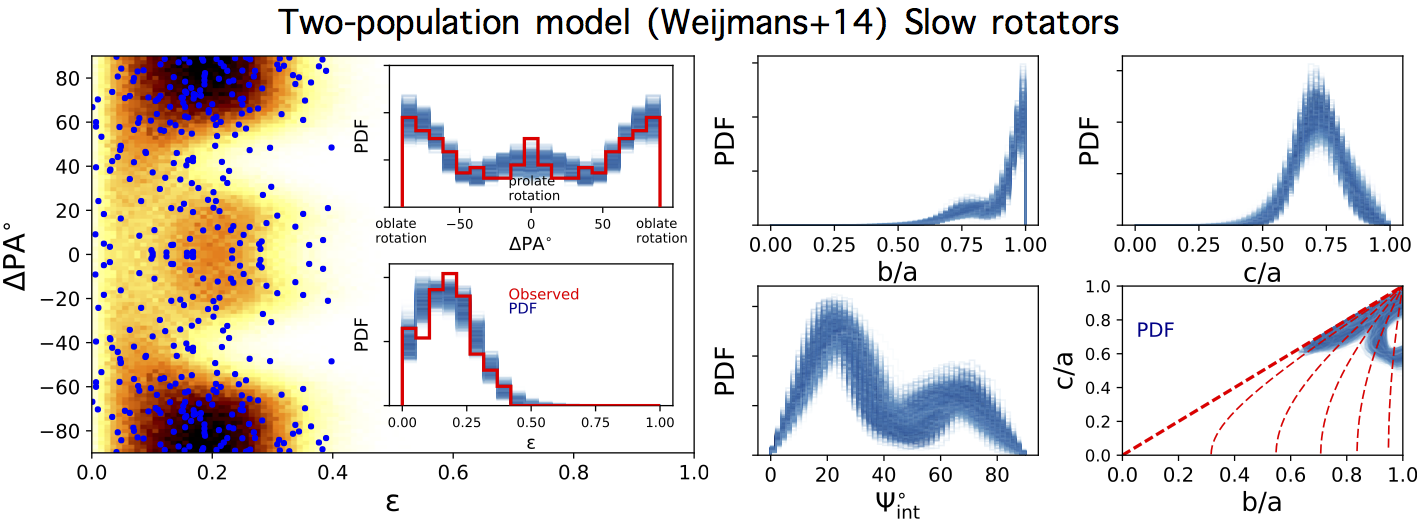}
\caption{Single-population (top) and two-population (bottom) models with the Weijmans+14 intrinsic
         misalignment for slow rotators. The other labels and legends are the same as Fig.~\ref{fig:single}.}
\label{fig:W14}
\end{figure*}

The model which maximizes the likelihood and the probability distribution function (PDF) from MCMC
for the single-population model with independent intrinsic misalignment is shown in Fig.~\ref{fig:single}.
From the probability distribution (blue histograms)
in the left panels, the model can produce 3 peaks (around $\Delta PA=0^{\circ}, \pm 90^{\circ}$),
which qualitatively agree with the observed data. However, the model fits the detailed structures
less well, e.g. the dip between $\sim 20^{\circ}$ and $\sim 40^{\circ}$, especially for non-regular
rotators. Furthermore, in the right panels, one can see that the model predicted axis ratios from different
samples do not agree with each other: galaxies in the slow rotator sample
prefer a triaxial-oblate shape with a flat intrinsic misalignment distribution while galaxies in the 
non-regular sample prefer a triaxial-prolate shape with flat intrinsic misalignment. 

Similar diagrams for the two-population model with independent intrinsic misalignment
are shown in Fig.~\ref{fig:double}. As can be seen from 
the probability distribution (blue histogram) in the left panels, the model can not
only give 3 peaks, but also match the amplitude of the peaks and other detailed structures 
relatively well. We also calculate the p-value from the KS statistic for the 1D distribution of $\Delta PA$.
We find that the p-value increases from $\sim0.8$ to $\sim1$ after we change to the two-population model.
From the probability distribution of the axis ratio distributions (blue histogram
and contours) in the right panels, the galaxy sample is dominated by a triaxial-oblate population,
while there are also some minor galaxy populations located in other regions. But due to large 
model degeneracy, it is difficult to constrain their axis ratios well. We note that in practice
it is difficult for the MCMC chain to converge under such large model degeneracy, unless it is
run with extremely long steps.
In general, the probability distributions of intrinsic axis ratios are similar for slow rotators
and non-regular rotators, despite the different selection criteria. 

In Fig.~\ref{fig:W14}, we show the results of single- and two-population models with Weijmans+14 intrinsic
misalignment for the slow rotator sample. Unlike the independent intrinsic misalignment model in
Fig.~\ref{fig:single}, the single-population model with Weijmans+14 intrinsic misalignment fails to reproduce
the observed peaks. In order to explain the observations, the model requires $\sim 70\%$ of the galaxies in
the sample have triaxial-oblate shape while the other $\sim 30\%$ have triaxial-prolate shape. 
Although model dependent, this gives us a hint that there might be a minor population in the slow rotators which
has triaxial-prolate shape.
The best-fit model parameters and their uncertainties from MCMC are listed in Table~1. 

In order to test whether our results are affected by outliers (e.g. galaxies with
no rotation or poor data qualities), we rerun our model including an outlier component with constant likelihood.
The results are still the same and the outlier fraction (free parameter in the fitting) is less than $5\%$
for all the models and samples. We also try a different axis ratio parametrization described
in \citet{Lambas1992}. The conclusions remain unchanged. 

\begin{table*}
    \tabcolsep=8.0pt
	\centering
	\caption{Best-fit (maximum likelihood) parameters for the models shown in Figs.~\ref{fig:single}, \ref{fig:double} and
    \ref{fig:W14}. The uncertainties are calculated as the differences between the $16$th and $50$th, $50$th and $84$th
    percentiles ($1\sigma$) of the 1D distribution of each parameter, respectively. We note that the best-fit parameters 
    can deviate from the $50$th percentile (i.e. the median), since the posterior distributions are irregular. SR and NR 
    represent slow rotators and non-regular rotators. SINGLE-INDEPENDENT represents the single-population model with
    independent intrinsic misalignment. $\rm TWO^1-INDEPENDENT$ and $\rm TWO^2-INDEPENDENT$ represent the two populations
    in the two-population model with independent intrinsic misalignment.
    SINGLE-W14, $\rm TWO-W14^1$ and $\rm TWO-W14^2$ represent the single-population
    model and the two components of the two-population model with the \citet{Weijmans2014} intrinsic misalignment.
    $\mu_{\psi_{\rm int}}$ and $\sigma_{\psi_{\rm int}}$ are in radians.}
	\label{tab:example_table}
    \footnotesize
	\begin{tabular}{lccccccc} 
		\hline
		model & $\mu_{\zeta}$ & $\sigma_{\zeta}$ & $\mu_{\eta}$ & $\sigma_{\eta}$ & $\mu_{\psi_{\rm int}}$ &  $\sigma_{\psi_{\rm int}}$ & $f_1$ \\
		\hline
         $\rm SR\quad SINGLE-INDEPENDENT$ & $0.99^{+0.01}_{-0.02}$ & $0.03^{+0.03}_{-0.01}$ & $0.74^{+0.02}_{-0.01}$ & $0.08^{+0.01}_{-0.01}$ & $0.14^{+0.58}_{-0.40}$ & $4.66^{+1.13}_{-1.26}$ &        -       \\
         $\rm NR\quad SINGLE-INDEPENDENT$ & $0.83^{+0.04}_{-0.06}$ & $0.13^{+0.03}_{-0.03}$ & $0.99^{+0.03}_{-0.04}$ & $0.04^{+0.04}_{-0.03}$ & $0.07^{+0.57}_{-0.44}$ & $0.87^{+1.28}_{-1.35}$ &        -       \\
         $\rm SR\quad TWO^1-INDEPENDENT$  & $0.99^{+0.05}_{-0.24}$ & $0.04^{+0.49}_{-0.06}$ & $0.78^{+0.12}_{-0.07}$ & $0.10^{+0.52}_{-0.03}$ & $0.35^{+0.63}_{-0.32}$ & $0.24^{+0.32}_{-0.64}$ & $0.55^{+0.45}_{-0.42}$ \\
         $\rm SR\quad TWO^2-INDEPENDENT$  & $0.86^{+0.06}_{-0.20}$ & $0.08^{+0.45}_{-0.06}$ & $0.82^{+0.14}_{-0.06}$ & $0.04^{+0.49}_{-0.03}$ & $1.55^{+0.74}_{-0.39}$ & $0.92^{+0.35}_{-0.57}$ &        -       \\
         $\rm NR\quad TWO^1-INDEPENDENT$  & $0.99^{+0.08}_{-0.21}$ & $0.04^{+0.49}_{-0.06}$ & $0.88^{+0.06}_{-0.25}$ & $0.13^{+0.50}_{-0.07}$ & $0.48^{+0.59}_{-0.41}$ & $0.10^{+0.39}_{-0.52}$ & $0.58^{+0.63}_{-0.13}$ \\
         $\rm NR\quad TWO^2-INDEPENDENT$  & $0.78^{+0.07}_{-0.19}$ & $0.09^{+0.40}_{-0.05}$ & $0.91^{+0.05}_{-0.21}$ & $0.05^{+0.51}_{-0.06}$ & $1.22^{+0.60}_{-0.38}$ & $1.09^{+0.37}_{-0.45}$ &        -       \\
         $\rm SR\quad SINGLE-W14$   & $0.99^{+0.01}_{-0.01}$ & $0.13^{+0.01}_{-0.01}$ & $0.82^{+0.03}_{-0.02}$ & $0.12^{+0.03}_{-0.02}$ &       -        &        -       &        -       \\
         $\rm SR\quad TWO-W14^1$ & $0.99^{+0.05}_{-0.19}$ & $0.05^{+0.27}_{-0.06}$ & $0.75^{+0.15}_{-0.10}$ & $0.08^{+0.03}_{-0.02}$ &       -        &        -       & $0.70^{+0.44}_{-0.47}$ \\
         $\rm SR\quad TWO-W14^2$ & $0.78^{+0.02}_{-0.18}$ & $0.08^{+0.09}_{-0.07}$ & $0.96^{+0.15}_{-0.08}$ & $0.05^{+0.09}_{-0.04}$ &       -        &        -       &        -       \\
		\hline
	\end{tabular}
\end{table*}

\section{Conclusions}
\label{sec:conclusion}
We study the distribution of apparent ellipticity and kinematic misalignment for slow rotators measured
in \citet{Graham2018}, based on a SDSS-IV DR14 MaNGA sample. The distribution shows clear structure
in the misalignment angle distribution: there are two peaks at both $0^{\circ}$ and $90^{\circ}$ 
misalignment (characteristic of oblate and prolate rotation respectively). By assuming Gaussian shapes
of the axis ratios and the intrinsic misalignment distribution, we invert the observed distribution to
obtain the intrinsic one using Monte Carlo simulations.

We find that our models with different assumptions clearly requires a dominant triaxial-oblate population,
beyond the known fact that slow rotators are weakly traxial.
If we assume the intrinsic shape and the intrinsic misalignment are independent, the constraints
on the intrinsic axis ratios are less strong. The observed distribution can be explained by a dominant
triaxial-oblate population (with some minor populations of different shapes). However, if a one-to-one
intrinsic shape-misalignment relation is assumed, the model requires two distinct galaxy populations 
(i.e. a dominant triaxial-oblate population and a minor triaxial-prolate population) in order to explain
the observations, suggesting a hint for a minor triaxial-prolate population.

The uncertainties in our study mainly come from the observational uncertainties of the misalignment 
angle and the ellipticity plus the intrinsic degeneracies in the inversion problem.
In the study, we measure the ellipticity, photometric and kinematic position angles around
the effective radius, and these values could vary with radius and cause uncertainties in the analysis.
In addition, galaxies with round shape and very slow rotation have large uncertainties in the misalignment
angle measurement. These uncertainties are difficult to estimate robustly and incorporate in the model.
Better observations and larger samples may help to reduce these uncertainties as well as the Poisson 
noise shown in Fig.~\ref{fig:obs}. The model degeneracies are intrinsic and difficult to remove. Our 
simulations show that the fitting results depend on the intrinsic misalignment model. This suggests
a better understanding of the intrinsic shape-misalignment relationship can be useful in the inversion.
A study in numerical simulation may be helpful.
The MaNGA sample we used is not volume limited, but has a flat stellar mass distribution \citep{Bundy2015}.
Since the galaxies we selected have a narrow mass range (between $2\times 10^{11} M_{\odot}$ 
and $\sim 10^{12} M_{\odot}$), this would only have a minor effect on the conclusions.

\acknowledgments

This work was supported by the National Science Foundation of China (Grant No. 11333003, 11390372
to SM). MC acknowledges support from a Royal Society University Research Fellowship.
We performed our computer runs on the Zen high performance computer cluster of the National
Astronomical Observatories, Chinese Academy of Sciences (NAOC), and the Venus server
at Tsinghua University.  This research made use of Marvin \citep{marvin}, a core Python package and
web framework for MaNGA data, developed by Brian Cherinka, Jos{\'e} S{\'a}nchez-Gallego,
and Brett Andrews (MaNGA Collaboration, 2017).

Funding for the Sloan Digital Sky Survey IV has been provided by
the Alfred P. Sloan Foundation, the U.S. Department of Energy Office of
Science, and the Participating Institutions. SDSS-IV acknowledges
support and resources from the Center for High-Performance Computing at
the University of Utah. The SDSS web site is www.sdss.org.

SDSS-IV is managed by the Astrophysical Research Consortium for the 
Participating Institutions of the SDSS Collaboration including the 
Brazilian Participation Group, the Carnegie Institution for Science, 
Carnegie Mellon University, the Chilean Participation Group, the French Participation Group, 
Harvard-Smithsonian Center for Astrophysics, 
Instituto de Astrof\'isica de Canarias, The Johns Hopkins University, 
Kavli Institute for the Physics and Mathematics of the Universe (IPMU) / 
University of Tokyo, Lawrence Berkeley National Laboratory, 
Leibniz Institut f\"ur Astrophysik Potsdam (AIP),  
Max-Planck-Institut f\"ur Astronomie (MPIA Heidelberg), 
Max-Planck-Institut f\"ur Astrophysik (MPA Garching), 
Max-Planck-Institut f\"ur Extraterrestrische Physik (MPE), 
National Astronomical Observatories of China, New Mexico State University, 
New York University, University of Notre Dame, 
Observat\'ario Nacional / MCTI, The Ohio State University, 
Pennsylvania State University, Shanghai Astronomical Observatory, 
United Kingdom Participation Group,
Universidad Nacional Aut\'onoma de M\'exico, University of Arizona, 
University of Colorado Boulder, University of Oxford, University of Portsmouth, 
University of Utah, University of Virginia, University of Washington, University of Wisconsin, 
Vanderbilt University, and Yale University.

\end{CJK*}
\end{document}